\documentclass[twocolumn]{aa}
\usepackage{natbib}\bibpunct{(}{)}{;}{a}{}{,}
\usepackage{graphicx}
\usepackage{txfonts}
\begin{document}

\title{Toward understanding the early stages of an impulsively
       accelerated coronal mass ejection}

 \subtitle{SECCHI observations}

 \author{S. Patsourakos \inst{1},
         A. Vourlidas   \inst{2}, \and 
         B. Kliem       \inst{3,4}}

 \institute{University of Ioannina, Department of Physics, 
            Section of Astrogeophysics, Ioannina, GR-45110, Greece\\
            \email{spatsour@cc.uoi.gr}
         \and
             Naval Research Laboratory, Space Science Division,
             Washington, DC 20375, USA \\ 
             \email{vourlidas@nrl.navy.mil} 
         \and 
             University College London, Mullard Space Science Laboratory,
             Holmbury St.\ Mary, Dorking, Surrey, UK
         \and 
             Universit\"{a}t Potsdam, Institut f\"{u}r Physik und Astronomie,
	     Potsdam, Germany\\
             \email{bkliem@uni-potsdam.de}}
  \date{}

  \abstract
  { The expanding magnetic flux in coronal mass ejections (CMEs) often
    forms a cavity. Studies of CME cavities have so far been limited to
    the pre-event configuration or to evolved CMEs at large heights, and
    to two-dimensional imaging data.
  }
   { Quantitative analysis of three-dimensional cavity evolution at
     CME onset can reveal information that is relevant to the genesis
     of the eruption.
   }
 { A spherical model is simultaneously fit to
     \textsl{Solar Terrestrial Relations Observatory (STEREO)}
     Extreme Ultraviolet Imager (EUVI) and Inner Coronagraph (COR1) data
     of an impulsively accelerated CME on 25 March 2008,
     which displays a well-defined extreme ultraviolet (EUV) and
     white-light cavity of nearly circular shape already at low heights
     $h\approx0.2R_\odot$. The center height $h(t)$ and radial
     expansion $r(t)$ of the cavity are obtained in the whole height
     range of the main acceleration. We interpret them as the axis
     height and as a quantity  proportional to the minor radius of a
     flux rope, respectively.
   }
   { The three-dimensional expansion of the CME exhibits two
    phases in the course of its main upward acceleration.
    From the first $h$ and $r$ data points, taken shortly after
    the onset of the main acceleration, the erupting flux
    shows an overexpansion compared to its rise, as expressed by
    the decrease of the aspect ratio from $\kappa=h/r\approx3$ to
    $\kappa\approx(1.5\mbox{--}2)$.
    This phase is approximately coincident with the
    impulsive rise of the acceleration and is followed by a phase of
    very gradual change of the aspect ratio (a nearly self-similar
    expansion) toward $\kappa\sim2.5$ at $h\sim10R_\odot$.
    The initial overexpansion of the CME cavity can be caused by flux
    conservation around a rising flux rope of decreasing axial current
    and by the addition of flux to a growing, or even newly forming,
    flux rope by magnetic reconnection. Further analysis will be
    required to decide which of these contributions is dominant.
    The data also suggest that the horizontal component of the impulsive
    cavity expansion (parallel to the solar surface) triggers the
    associated EUV wave, which subsequently detaches from the CME volume.
    }
   {}

\authorrunning{Patsourakos et al.}
\titlerunning{Understanding the early expansion of an impulsive CME}

   \maketitle

\section{Introduction}\label{intro}

 
\begin{figure*}
\includegraphics[scale=.85]{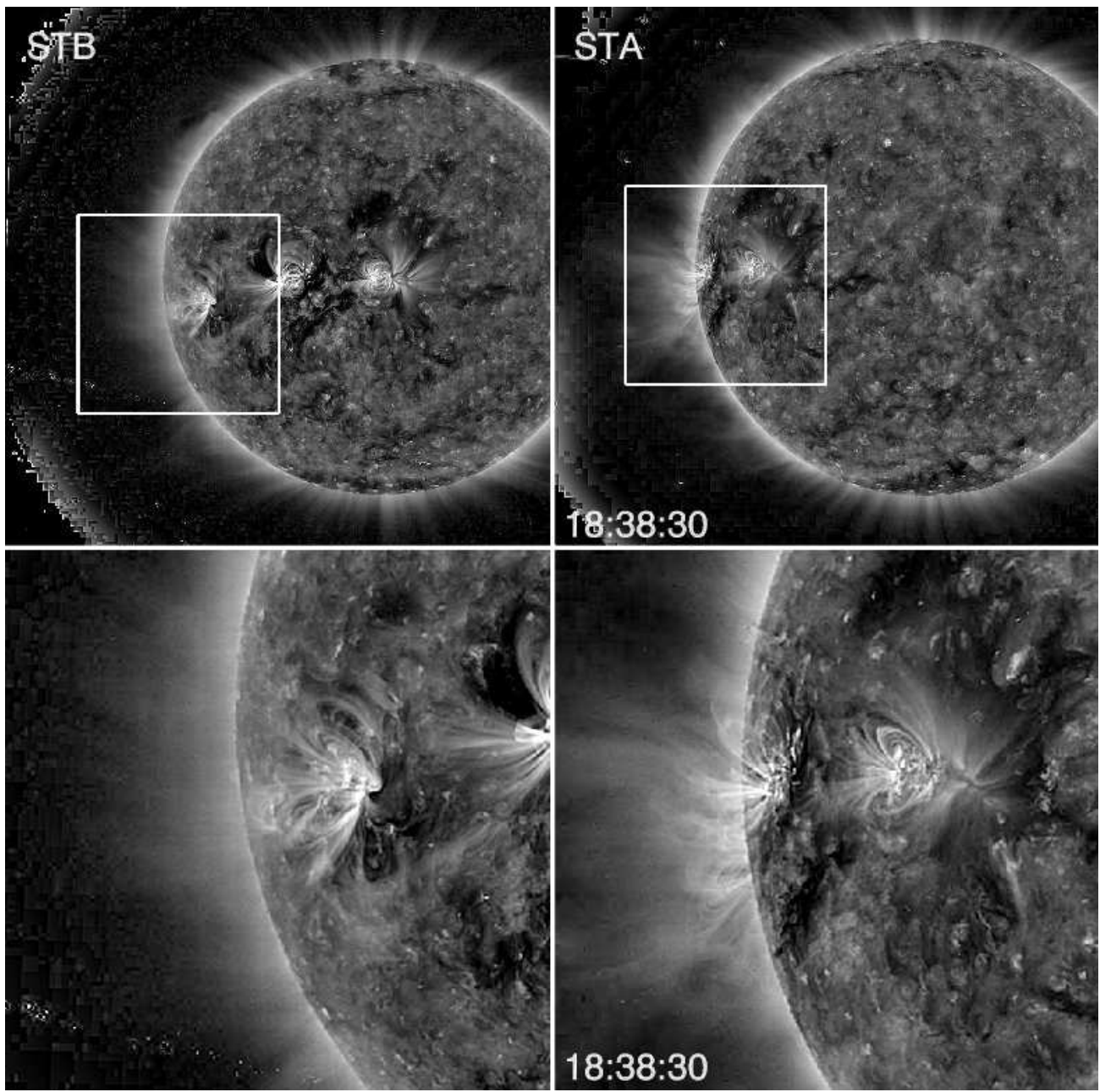}
\caption{Overview of the pre-eruption configuration at 18:38:30~UT on
  25-Mar-2008 in STB (left column) and STA (right column).
  Wavelet-enhanced EUVI 171~\AA\ images are shown. The images
  in the bottom row are enlargements of the erupting active region,
  which is marked by boxes in the upper panels. The temporal evolution
  is shown in the movie in the on-line edition (movie1.mpg).
  Solar north is on top.}
\label{cont1}
\end{figure*}

The genesis of coronal mass ejections (CMEs) has been the subject of
intense study and significant controversy since their discovery
\cite[e.g.,][]{2000JGR...10523153F, 2001AGUGM.125..143K,
2005ARA&A..43..103Z, 2006SSRv..123..251F, 2006SSRv..123...57M}.
Fast CMEs tend to reveal the main physical effects at work most clearly,
due to their high energy release rate and large total energy. However,
these events exhibit impulsive acceleration low in the corona
\citep{MacQueen&Fisher1983, 1999JGR...10424739S, Vrsnak2001,
2006ApJ...649.1100Z}, so that they are most difficult to resolve.
A number of case studies over the previous solar cycle have
shown that the ejected plasma can be accelerated within a
couple of minutes and at heights as low as $0.05R_\odot$
(see, e.g., Dere et al. 1997, 1999; Zhang et al. 2001; Gallagher et al.
2003; Williams et al. 2005; Schrijver et al. 2008).
Also, a close temporal relationship was found between the acceleration
of fast CMEs and the X-ray flux of the associated flare
\citep{2001ApJ...559..452Z, Neupert&al2001, 2007SoPh..241...99M,
2008ApJ...673L..95T}.
Significant observational progress is now enabled by the
combination of high spatial resolution, high cadence, and complete
height coverage provided by the Sun Earth Connection Coronal and Heliospheric
Investigation (SECCHI) instruments aboard the
\textsl{Solar Terrestrial Relations Observatory (STEREO)},
in addition to the imaging from two or three viewpoints.


Some CMEs reveal an emission void in the EUV, or a cavity in white-light
images, in the source volume of the eruption prior to or early in
the main acceleration phase (Dere et al. 1997; Plunkett et al. 1997,
2000; Maia et al. 1999; Gibson et al. 2006). Such a void is usually
interpreted as the low-corona counterpart of the cavity seen further out
in coronagraph images of ``three-part'' CMEs. Those large-scale cavities
can be modeled as the cross section of a flux rope (but may be larger
than the actual cross section; Chen et al. 1997;
Wood et al. 1999; Vourlidas et al. 2000;  Krall et al. 2001; Thernisien
et al. 2006; Subramanian \& Vourlidas 2007). They are now widely
considered to be the signature of an expanding flux rope. This view was
recently confirmed by stereoscopic observations (Thernisien et al. 2009;
Wood \& Howard 2009). Hence, the flux rope interpretation for the
EUV void, or analogous white-light cavity at low heights, has strong
support. However, the interpretation as a sheared arcade, or as a
sheared arcade that contains a flux rope in its center,
is not excluded for such structures observed prior to an eruption.


Important questions pertaining to fast CMEs (and CMEs in general)
include the following.
Can the EUV or white-light cavities seen at low heights early in
the evolution of CMEs be interpreted as flux ropes?
What is their 3D shape?
What is their evolution; in particular, do they expand self-similarly?
Is their evolution relevant for EUV waves?


Obtaining answers to these questions would allow
for significant progress in our understanding of CMEs. 
For example, establishing a relationship between EUV cavities and
the cavities seen in white-light CMEs would indicate the presence of a
flux rope at lower heights than accessible to coronagraph observations.
Determining the 3D structure of EUV cavities should allow for
further insight on whether they represent early instances
of a flux rope or merely reflect how the evolving CME perturbs the
ambient field, i.e., whether the cavity as a whole or only its inner
part is the volume of a flux rope.
The lateral evolution of CMEs in their early phase is largely unexplored.
This subject provides a
new diagnostic of the physical processes that operate during CME onsets,
relevant to the importance of ideal vs. non-ideal effects.
Finally, constraining the erupting volume at high cadence by 3D
observations and modeling allows to address the question whether the
expanding CME acts as an impulsive trigger or as a continuous driver of
the EUV wave which is often associated with fast eruptions.

In the present paper we utilize the unique capabilities provided by the
\textsl{STEREO} instruments---stereoscopic imaging of the outer \emph{and}
inner corona at high spatial resolution and cadence---to
derive quantitative information relevant to the evolution of a flux
rope during the nascent stages of a fast CME.
The \textsl{STEREO} mission \citep{Kaiser&al2008} was launched in late
2006. It consists of two almost identical spacecraft orbiting the Sun
at different distances with one trailing the Earth's orbit (\textsl{Behind}
spacecraft; STB)
and the other leading it (\textsl{Ahead} spacecraft; STA). The angular
separation between the two spacecraft changes
at a rate of about 45~degrees per year which allows for  truly multi-viewpoint
observations of the Sun and heliosphere.


\begin{figure}
\includegraphics[scale=.65]{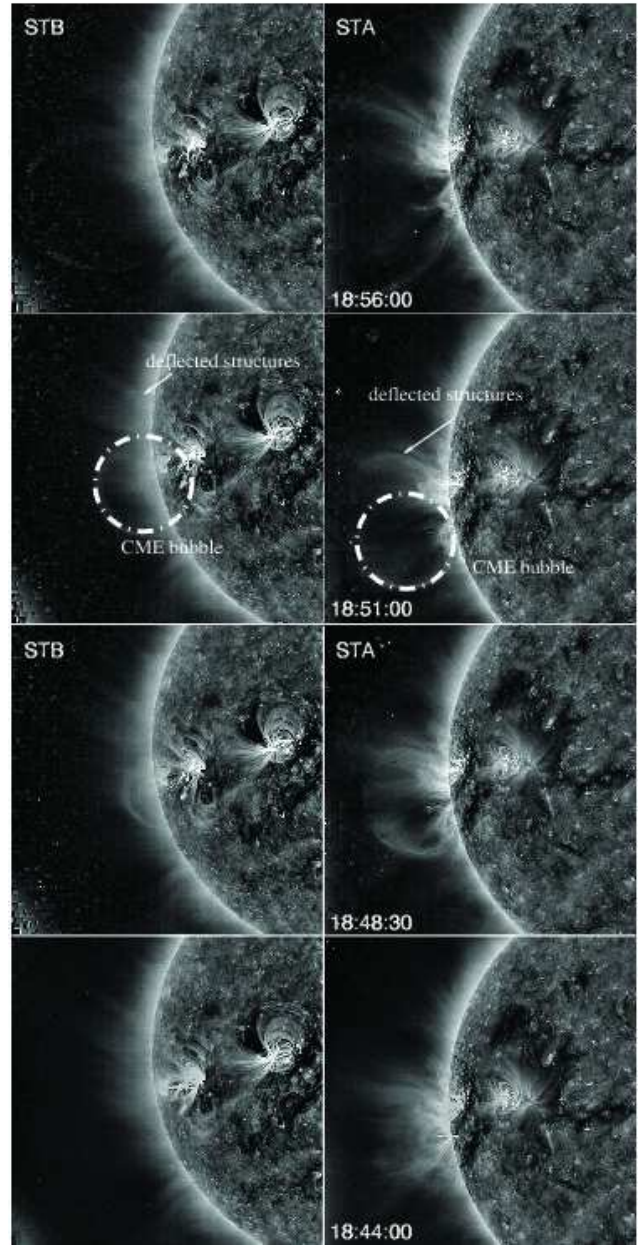}
\caption{Selected EUVI 171~\AA\  snapshots during the eruption sequence.
  Left column: STB, right column: STA. Time increases from bottom to top.
  The dash-dotted cicles in the 18:51~UT images show our identification of
  the EUV cavity (``bubble''). Also marked are deflected structures,
  which can be easily misidentified as part of the erupting bubble if no
  proper 3D analysis is performed (Sect.~\ref{proper}). The temporal 
  evolution is shown in the movie in the on-line edition (movie2.mpg).}
\label{cont2}
\end{figure}

A structure reminiscent of a void became a very conspicuous rapidly
evolving feature of a CME on 25 March 2008.
The structure quickly developed the shape of a bubble,
and we will refer to it as a ``bubble'' in the following. The near-limb
location of the eruption, combined with the large continuous height
coverage and high cadence,
revealed the evolution of the bubble into the CME cavity unambiguously.
However, only the proper interpretation of the information from both
viewpoints allows us to relate the bubble correctly to the whole
volume of expanding magnetic flux. The  geometrical parameters of the
bubble are then derived
by fitting a three-dimensional model to its images.
This enables an in-depth quantitative study of its evolution
free of projection effects and free of confusion between the bubble and other
structures aligned along the line of sight. In particular, we obtain the
aspect ratio, interpreted as being proportional to the
ratio of axis height and minor radius of a flux rope, at the earliest
stages of a CME cavity. These measurements can address the important
questions when a flux rope forms in CMEs and what determines its initial
evolution.
We also consider the association between the CME and the accompanying
flare with regard to the role of reconnection, taking advantage of the
relatively high cadence of the data. Finally, the associated EUV wave
is briefly discussed. We relate its triggering to the expansion of the
cavity.

This paper is organized as follows. An overview of the observations is
given in Sect.~\ref{over}. Section~\ref{proper} discusses in detail
the two-viewpoint high-cadence observations of the bubble and its
proper definition. Sect.~\ref{bubble} presents the 3D geometrical
modeling of the bubble, the results of which are analyzed and
discussed in Sects.~\ref{timing} and \ref{interpretation}, respectively.
The EUV wave is considered in Sect.~\ref{s:wave},
and Sect.~\ref{concl} summarizes our findings.

\begin{figure*}
\sidecaption
\includegraphics[width=12cm]{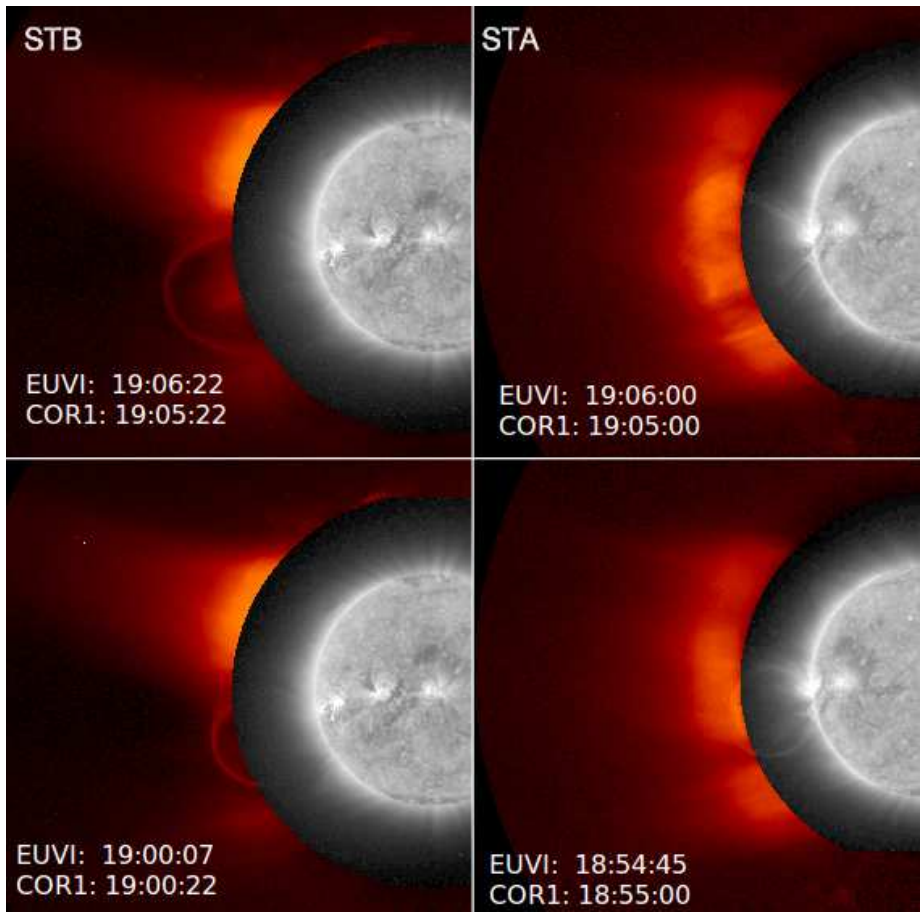}
\caption{Composite of EUVI 171~\AA\ and COR1 white-light images from STA
  and STB showing the emergence of the EUVI bubble into the COR1 field
  of view, where it becomes the cavity of a typical three-part CME.} 
\label{fest}
\end{figure*}

\section{Overview of the observations}\label{over}

We use observations from the Extreme Ultraviolet Imaging Telescope
(EUVI; Wuelser et al. 2004) and the inner white-light coronagraph
(COR1; Thompson et al. 2003) of the
SECCHI instrument suite (Howard et al. 2008)
on-board the two \textsl{STEREO} spacecraft.  EUVI is a full disk imager
with a field of view extending to 1.7$R_{\odot}$. We mainly use data from the
171\AA\ channel (hereafter 171) because these have the highest cadence (75~s
in STA, 150~s in STB). The EUVI data are supplemented with total
brightness images from COR1, which observes the white-light corona from
1.5 to 4$R_{\odot}$ with a 10 minute cadence.
Image pairs are taken with a time difference (around 20 s 
for the day of our observations) between
STA and STB which compensates for the different light transit times
to each spacecraft; therefore such data are synchronized on the Sun.

A CME-flare event took place in NOAA active region (AR) 10989 on 25
March 2008 after about 18:30~UT. The center of the region was located
at coordinates $10^\circ$E $80^\circ$N  as seen from the Earth
during the period of the event. The \textsl{STEREO} spacecraft had a
separation of $\approx\!47^\circ$, with STB viewing the active region
$\approx\!57^\circ$ east of central meridian and STA viewing it slightly
occulted near the east limb (Fig.~\ref{cont1}).

This event was preceded by weaker activity on the far side of the
neighboring AR~10988 located $30^\circ$ west of AR~10989. A filament
activation started near 18:20~UT, accompanied by a
small, B1.8 X-ray class flare which peaked at 18:29~UT, which could be
considered as a possible trigger for our event based solely on the timing.
However,
the EUVI data show no evidence of interaction between the two active
regions after that early compact flare (no dimmings, no waves, no
dissapearing loops, etc). In fact, the flows and flaring continued in
AR~10988 after our event again without any direct evidence of
interaction between the activities in the two regions.

The first indication of an eruption in AR~10989 was observed as a
slow rise of loops in the center of the region, starting
at $\approx$\,18:30~UT. This was followed by an explosive expansion
forming a large bubble  after $\approx$\,18:41~UT. Tbe bubble eventually
became part of a fast white-light CME which reached a velocity of
$\approx\!1100~\mathrm{km\,{s}^{-1}}$ in the outer corona (2.5--15$R_{\odot}$)
as measured in Thernisien et al. (2009) and Temmer et al. (2010). The associated flare
was an M1.7 \textsl{GOES} class starting at $\approx$\,18:36~UT and peaking at
$\approx$\,18:56~UT. The soft X-ray flux remained above pre-event levels for
several hours after the event.

We have posted two EUVI 171 movies in the online version of the paper
which show the evolution of the event from both spacecraft
simultaneously. The first movie (movie1.mpg) shows a large-scale view
of the event and the second movie (movie2.mpg) zooms in over the
erupting bubble. Selected snapshots are shown in Figs.~\ref{cont1} and
\ref{cont2}. Besides the standard image reduction for SECCHI data,
the 171 images were also processed using wavelets to enhance their
contrast (Stenborg et al. 2008).

A snapshot of the pre-event configuration is given in
Fig.~\ref{cont1}.  The source region is dominated by low-lying loops,
many of them inclined, as seen by STB. On the other hand, the STA images show
some more extended and diffuse coronal structures around the active region
besides the low-lying loops. 

These data do not provide any definite evidence for the existence
of an EUV cavity prior to the eruption. This may be due to the
inclination of the polarity inversion line in the center of the
region by $\sim\!45^\circ$ to the east-west direction. A faint
indication of reduced intensity can be found in a narrow vertical
area at the southern edge of the active region (see the zoomed STA
view). The polarity inversion line bends slightly eastward in this
area. However, these effects are far too weak to be conclusive.
Note finally that an inspection of the High Altitude Observatory's
Mk4 Coronameter images taken on
the day of the event did not yield any evidence for the
existence of a cavity before the eruption at heights $h>1.14R_\odot$.

The slow rise of the low-lying loops in the center of the active region
changes into a rapidly accelerating evolution between about
18:35 and 18:41~UT,
which is most obvious in the near-limb view of STA. A bubble
forms at about this time and expands rapidly as it rises. It
develops a relatively sharp rim (Fig.~\ref{cont2}). The strongest visual
changes occur between about 18:44:45 and 18:48:30~UT when the rim has
developed and expands very rapidly. The rise of the bubble accelerates
over a slightly longer time scale, so that the images indicate a
sequence of fast expansion of the erupting volume, followed by fast rise.
This will be quantified by the analysis in Sects.~\ref{bubble} and
\ref{timing}.

Intense brightenings can be seen in the active region core, marking the
developing flare. A prominence also erupts, forming the peaked shape
of a kinking flux rope \cite[e.g.,][]{Torok&Kliem2005} before it fades.
The prominence  material lies in the bottom part of the bubble in those
frames that show both structures (see the panel at 18:48:30~UT in
Fig.~\ref{cont2}). Striated structures, resembling loops seen edge-on,
can be seen in the interior of the bubble.

\begin{figure*}
\sidecaption
\includegraphics[width=12cm]{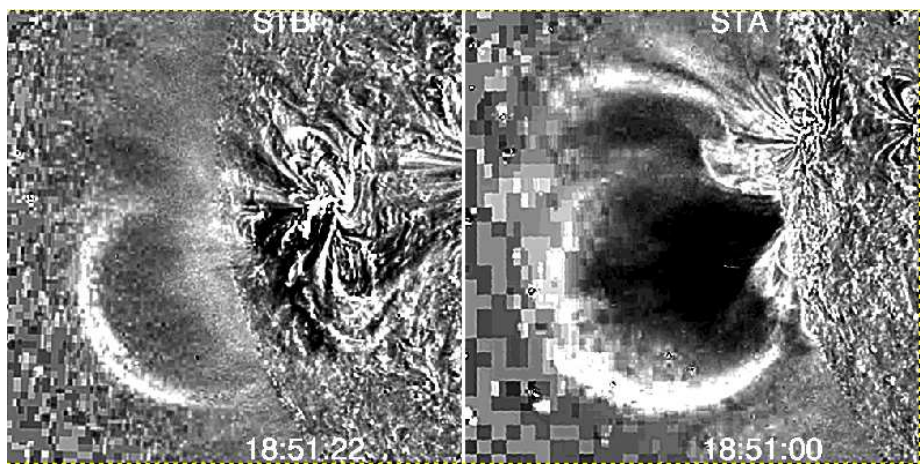}
\caption{Base-ratio images for representative snapshots during the event. 
  Each EUVI 171~\AA\ image is divided by the corresponding pre-event
  image taken at 18:39:45~UT. The resulting ratios are scaled into
  the range 0.2--1.8 for each frame. The pixel size is not rescaled, so
  that the solar radius differs slightly, being smaller by
  $\approx\!5\,\%$ for STB. The temporal evolution is shown in the movies 
  in the on-line edition (movie3.mpg, movie4.mpg).}
\label{ratio}
\end{figure*}

The STA limb view also shows that the expanding
bubble pushes aside nearby coronal structures. As time advances,
deflections of ambient coronal structures are visible at progressively larger
distances from the bubble revealing the propagation of a wave.  Indeed,
an EUV wave can be seen traveling away from the bubble in 195, 284 and 
171 channel observations from both STA and STB spacecraft (see
Sect.~\ref{s:wave}).

The bubble continues to grow and
to propagate outward, its front exiting the EUVI field of view at 18:55~UT.
A large intensity depletion (dimming) is seen in its wake marking the
mass evacuation associated with the eruption.
Finally, we note the existence of
concave-upward and striated structures at the bottom of the erupting bubble 
(see, e.g., the STA data after 18:54~UT and the STB data between
18:57:15 and 18:59~UT).  Such features are
commonly interpreted as an indication of a flux rope structure
(e.g., Plunkett et al. 2000).

Eventually the bubble becomes the white-light CME. This is apparent in
the composite of nearly-simultaneous EUVI 171 and COR1 images in
Fig.~\ref{fest}.  The FESTIVAL software (Auch{\`e}re et al. 2008)
was used to generate these images.  The erupting bubble evolves into
the CME cavity once it enters the COR1 field of view.
Note, for instance, the correspondences between the EUVI
bubble and CME front in STA at 18:55~UT and between the bubble and
CME flanks in STB at 19:05~UT. It is well established that the
cavity of white-light CMEs is  a signature of an erupting flux rope
(see the Introduction); in particular, a flux rope model was
successfully fit to the stereoscopic COR2 data of this event at
heliocentric distances $>2.5R_\odot$ (Thernisien et al. 2009). Hence, we can
infer,  with high likelihood, that the EUV bubble is the low coronal
signature of the CME flux rope.

\section{Proper definition of the erupting bubble}\label{proper}

From both spacecraft, the bubble is seen as a rim of enhanced brightness
enclosing a dimming area (Figs.~\ref{cont2} and \ref{fest}). Ratio or
difference images bring out such structure much clearer. Animations of
base-ratio 171 images from STA and STB (with each image of the sequence
divided by a base pre-event image) are posted in the online edition
of the paper as movie3.mpg and movie4.mpg, respectively, and two
nearly simultaneous frames from these animations are shown in 
Fig.~\ref{ratio}. These confirm the basic structure of a rim, which must
consist of coronal plasma accumulated  at the surface of the expanding
magnetic flux.


Since the optically thin EUV emission results from the integration of 
all column density along the line of sight, there are two
possibilities to produce this structure in the brightness distribution.
(1) A relatively unstructured density distribution at the
surface of the expanding volume, leading to enhanced column density at
the projected edge in the plane of the sky where the observer's line of
sight is tangential. Different perspectives map different parts of the
surface onto the corresponding plane of the sky.
(2) Loop-shaped enhancements of the
surface density due to overlying loops which have been swept up. In this
case, the same traces on the surface are mapped onto the plane of the
sky for any perspective of the observer, and the true edge of the
expanding volume may extend beyond the rim in the images. The data favor
the first interpretation because the bubble rim in the EUVI images
evolves continuously into the rim of the CME cavity in the COR1 and COR2
images, indicating that it is the true edge of the expanding volume, and
because the rim is close to a circle from the perspectives of both STA
and STB. The latter property is incompatible with an essentially planar
loop-shaped density enhancement, which would appear notably elongated
(elliptical) in at least one of the views, due to the considerable
angular separation of the spacecraft.


Comparing the size and location of the rim between STA and STB, one
finds that a mismatch gradually develops from about 18:46~UT onwards.
While the southern part of the rim retains a good correspondence
between the STA and STB images throughout the event, the northern part
of the rim for STA expands beyond the rim for STB.

This feature was also noted by
\citet{2009AnGeo..27.3275A} and is very clearly seen in his Fig.~3,
where circular fits to the evolving rim for STA are overplotted on the
corresponding STB base-difference images. Since the northern part of the STA
rim extends beyond the STB rim in a direction perpendicular to the
direction of the angular separation between the spacecraft (which is
nearly in the ecliptic), the two rims cannot map the same edge of the
expanding volume. In order to develop an interpretation for this
feature, let us  consider the base ratio images in more detail.


The STA image in Fig.~\ref{ratio} shows that the deepest dimming occurs
in a direction inclined from radial toward the south. This is also the
direction of ascent of the bubble in the STB images and the direction of
ascent of the prominence. The deepest dimming defines the core of the
eruption, and in the STA ratio images it is enclosed on the northern
side by a second, less bright rim, which can be discerned from 18:47:15 to
18:54~UT. This part of the eruption has approximately the same size
for STA and STB and appears consistent with a nearly spherical object
throughout the EUVI field of view. A sphere does indeed yield a
consistent fit to the STA \emph{and} STB EUVI images of this
structure; the fit remains consistent even out to the first two COR1
image pairs (Sect.~\ref{bubble}). Hence, we adopt this bubble-shaped
volume of deepest dimming as the core part of the eruption, to be
modeled in this paper by 3D fitting to its rim. This choice is supported
\emph{a posteriori} by the timing of the resulting upward acceleration
profile of the bubble, whose onset coincides with the first signs of the
associated flare's soft X-ray emission (Sect.~\ref{timing}).

Our definition of the bubble also leads to a plausible
interpretation for the different expansions seen by STA and STB after
about 18:46~UT. We suggest that the expanding volume evolved from a
nearly spherical shape (as long as the two rims  are of nearly
equal size and nearly cospatial in projection), to the shape of a torus
section with conical legs (similar to a croissant), which was
successfully fitted to the stereoscopic images at later stages of the
event in the height range of COR2 \citep{2009SoPh..256..111T}.
This transition from a sphere to a torus may be associated to a `zipper'
effect, when reconnection progresses not only upwards but also in a
direction \emph{along} the neutral line at the base of the eruption. The
STB data indeed show that the post-eruption loop arcade grows
considerably along the neutral line in the south east direction after
18:47~UT. Here we assume that the magnetic axis
of the structure rotated by $\sim\!45^\circ$ in the clockwise direction
from the original alignment with the polarity inversion line to become
approximately aligned with the east-west direction. This orientation of
the axis is indicated by the depth of the dimming seen by STA
(Fig.~\ref{ratio}) and by the shape of the white-light cavity at greater
heights (see the STB COR1 image at 19:05~UT in Fig.~\ref{fest}). The
rotation is consistent with the observed kinking of the embedded
filament if the field is right-handed \citep{Green&al2007}, which, in
turn, is consistent with the location of the active region in the
southern hemisphere.

We note that the dimming seen by STB above the limb is much weaker than
the STA dimming (Fig.~\ref{ratio}). This suggests that the magnetic axis
of the erupting flux was much closer aligned to the line of sight for
STA (which is plausible due to the near-limb location of the active
region for STA), so that STB was looking at the magnetic axis of the
erupting flux from an elevated perspective. This implies that much of
the STB dimming should lie on disk, as observed, and that the STB rim
maps the edge of the bubble somewhere between the top and the rear side
of the bubble.

\begin{figure*}
\includegraphics[width=.85\textwidth,angle=90]{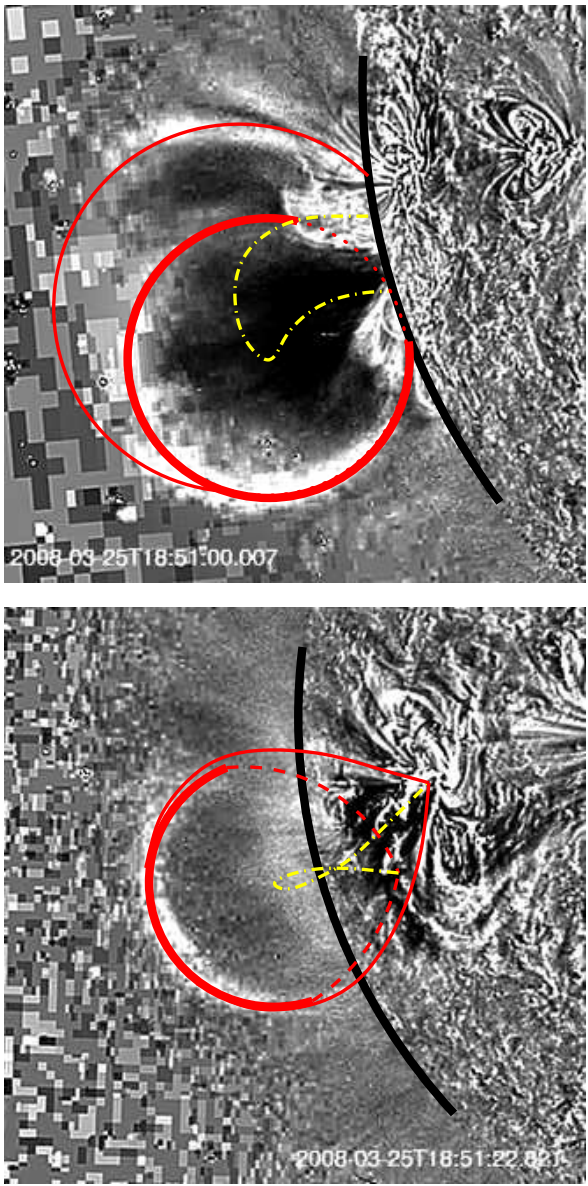}
\caption{Outlines of the envisioned croissant-shaped flux rope with
  a bubble-like enhancement in its center;
  \textsl{left:} STB view, \textsl{right:} STA view.
  Solid red lines represent edges of the erupting flux which become
  visible as rims in the EUVI images. Thick arcs indicate the bubble
  (edge of middle cross section for STA and edge of cross section in
  rear part for STB). Thin arcs indicate the croissant-shaped extensions
  of the bubble to the photospheric sources of the erupting flux. Dashed
  and dotted red lines indicate edges which are not visible (because they
  nearly coincide with other edges, are occulted, or are at different
  temperature). Dash-dotted yellow lines
  indicate the magnetic axis of the structure.
  See Sect.~\ref{proper} for further explanation.}
\label{f:sketch}
\end{figure*}


\begin{figure*}
\includegraphics[scale=.70]{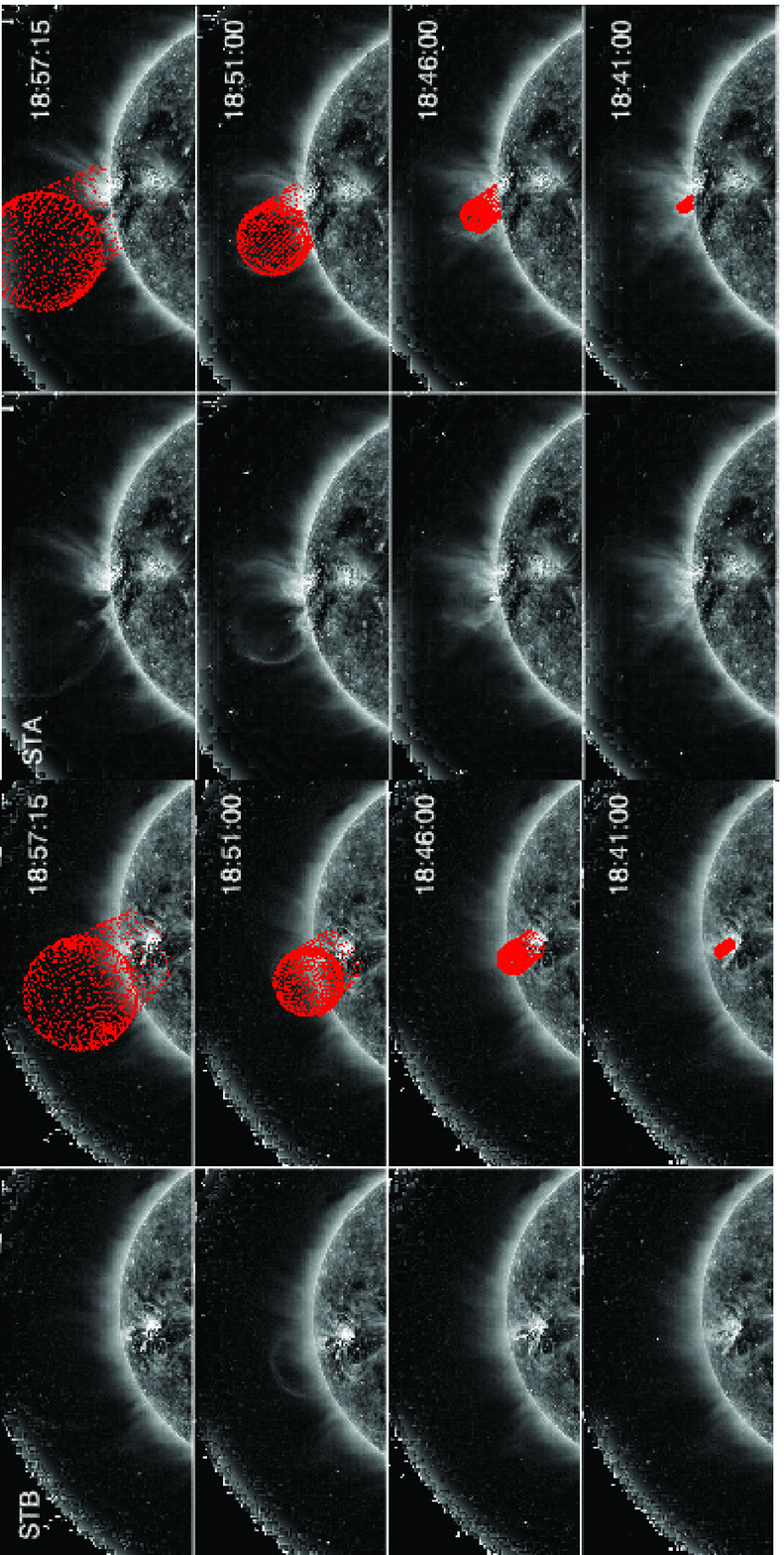}
\caption{Representative fits of the bubble by our 3D ice-cream cone
  model in EUVI 171~\AA\ data from STA and STB.  
  Each STA and STB image is shown twice: once without and
  once with the 3D model (red wireframe) overplotted.} 
\label{fitb}
\end{figure*}

A spherical volume can, of course, only be an approximation to the major
part of the erupting flux; the field lines must still connect to the
source areas of the flux in the photosphere. As the bubble radius
exceeds the size of the active region, which occurs already at low
heights, the source areas stay much smaller than the rim in the EUVI
images, so the extensions of the bubble approximate short cones. This
leads to a nearly spherical inverse teardrop shape in projection onto
the plane of the sky, fully compatible with the STA and STB images
during the first minutes of the expanding bubble. A croissant shape
results when the bubble begins to extend along the magnetic axis of the
erupting flux in the course of its further rise, to eventually approach
a partial toroidal shape of the upper part. For this evolving geometry, a
smaller rim can be associated with the cross section of the bubble or
middle part of the croissant (the minor radius of the developing torus),
while a progressively larger rim can be associated with the upper edge
of the growing croissant (the major radius of the developing torus) if
it is seen slightly from the side. Here the initially nonradial
propagation of the bubble is essential. Since the erupting field must
remain rooted in the active region, both the front and the rear
extensions of the bubble to the tips of the croissant are arranged
obliquely, in the northwestward direction. Hence, they are seen slightly
from the side (and probably slightly from below) by STA and
seen significantly from above by STB.

This geometry is sketched in Fig.~\ref{f:sketch}. The STA view on the
right-hand side is approximately aligned with the magnetic axis of the
flux rope in the middle (bubble) part of the croissant (looking at
the axis slightly from below). STA resolves both aspects of the
erupting structure: the whole edge of the bubble (thick red lines)
and the upper edge of the croissant-shaped frontal extension from the
bubble toward the northwest footpoint
(larger thin red line), which lies behind the STA limb. The lower edge of the
front extension (not included in the plot) nearly coincides with the
left (southern) part of the bubble edge in projection and is not
detected as a separate structure. The rear extension of the croissant
(not included in the plot) is nearly cospatial with the front extension
in projection, so that it does not produce separate rims.

The STB view on the left-hand side is approximately aligned along the
east-west direction with the magnetic axis of the  flux rope in the
middle (bubble) part, but elevated, i.e., looking down at the bubble.
STB sees the edge of the bubble in its upper-rear part (thick red line).
The images also show, albeit only very faintly, the left and right edges
of the front croissant-shaped extension from the bubble to the northwest
footpoint (on disk for STB). The rear  croissant-shaped extension
(dashed red lines) lies
essentially under the bubble and is not very different from the front
extension in projection, and thus does not produce separate rims.

The suggested \emph{evolving} geometry of the erupting flux explains
in a natural manner the persistence of the dimming between the
small and large rims in STA and that it is weaker than the core dimming, and
why the CME bubble shape is much less defined and more extended in the
COR1-A images (Figs.~\ref{fest} and \ref{fitcor1}). In a nutshell,
the core of the eruption in COR1-B (e.g., at 19:05~UT in
Fig.~\ref{fitcor1}) is part of the distorted structure in
COR1-A. The actual bubble is visible, albeit very faintly in the
background, and was the basis for the 3D fits shown in the figure.

Although our 3D modeling in Sect.~\ref{bubble} focuses exclusively on
the core of the eruption, it is important to understand how the major
bright and dimming structures are related between STA and STB, as well
as between the EUVI and the COR1--COR2 height ranges. Only a consistent
picture between STA and STB justifies the use of 3D modeling as a method
superior to individual 2D modeling of the STA and STB data. As we will
see in Sect.~\ref{timing}, the implications of the 3D modeling differ
drastically from a previous modeling \citep{2009AnGeo..27.3275A} which
focused exclusively on the large rim seen by STA after it began to
differ from the rim seen by STB, assuming that this rim maps the
edge of a spherical bubble.

The difference between Aschwanden's and our fitting is further amplified
by the different judgment of the ``deflected structures'' mentioned in
Sect.~\ref{over} and marked in Fig.~\ref{cont2}. In the STA images,
these structures are bent to the side when the erupting flux has expanded
sufficiently in horizontal direction up to their position. While some of
these originally nearly radial structures then coincide in projection
with the northern rim, others stay slightly in front of the rim. The
pre-event STB view (Fig.~\ref{cont1})
shows that these structures are not physically connected to the
bubble; they are rooted in a different part of the active region, at its
periphery. By fitting a circle to the deflected structures ahead of the
north rim in some of the STA images, Aschwanden obtained modeled bubble
outlines that are even larger than the rim and extend clearly beyond the
range of coronal off-limb dimmings associated with the CME.

In the subsequent 3D modeling we focus on the core of the eruption,
which is imaged by the 171 channel of both spacecraft. We consider the
core (the ``bubble'') to be defined by the deepest dimming for STA
and the rim enclosing it, and by the rim seen by STB.

\section{3D geometrical modeling of the erupting bubble}\label{bubble}

The considerations of the previous section lead us to adopt the
spherical geometry for the core of the CME as long as we can discern
rims of similar size and location which define a (not necessarily
complete) circular shape to sufficient accuracy in the 171 STA
\emph{and} STB images. This
is the case throughout the EUVI field of view, as well as for the first
two COR1 image pairs. It is clear that the assumption of spherical shape
becomes progressively more approximate as the erupting flux reaches
greater heights, since the second STA rim, which develops after about
18:46~UT, can be interpreted as a signature of a more elongated,
croissant-like shape and since a ``developed'' flux-rope shape gave a
successful 3D fit to the data in the COR2 field of view
\citep{2009SoPh..256..111T}.
 
The Graduated Cylindrical Shell model employed by
\citet{2009SoPh..256..111T} consists of a partial torus with radially
aligned conical legs.
It does not only fit the 3D morphology of evolved CMEs in the height
range imaged by COR1 and COR2, but also fits shocks (Ontiveros \&
Vourlidas 2009) and EUV waves (Patsourakos \& Vourlidas 2009).
For our purpose, we simplify the model by adopting coinciding legs,
which collapses the toroidal section into a sphere
(see Figs.~1 and ~2 in Thernisien et al. 2009). This geometry is
known in CME modeling as an ``ice-cream cone'' model.

The model is constructed with the following free parameters: position of
the leg on the solar surface (longitude, latitude), north-south tilt of
the leg with respect to the radial direction, and,
finally, distance to the  front and radius of the spherical shell.
The distance to the midpoint of the spherical shell, $h$, approximates the
axis height of the developing flux rope, while the radius, $r$, is assumed
to be proportional to the rope's minor radius.
Note that the rim of the bubble and the edge of the supposed flux
rope do not necessarily have to coincide. In Sect.~\ref{interpretation} we
discuss two possible interpretations of  rapid bubble expansion; one of
them identifies the bubble with the flux rope, while the other
identifies the rim with a flux surface in the ambient field.

\begin{figure*}
\includegraphics[scale=0.85]{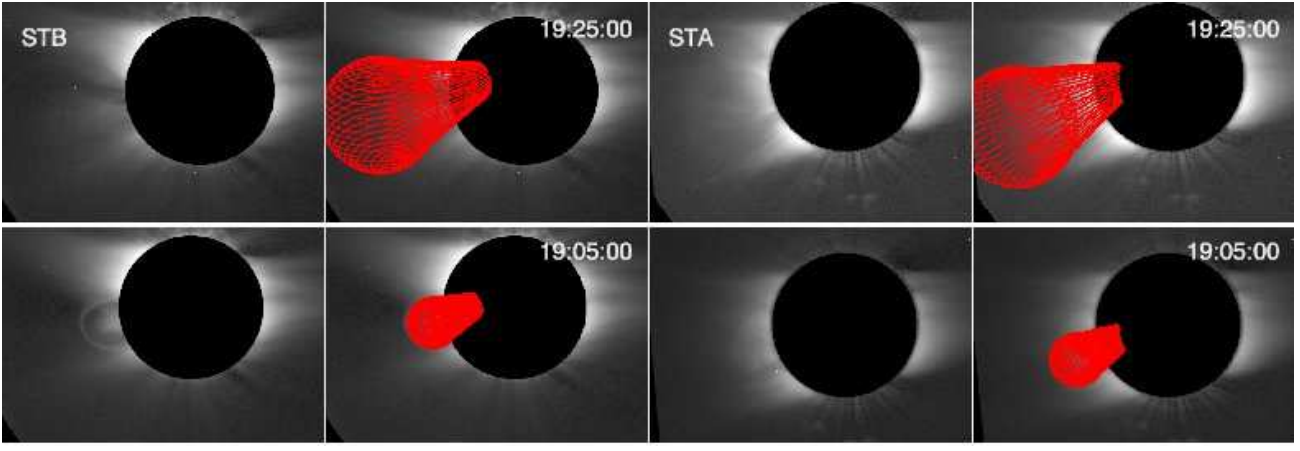}
\caption{Similar to Fig 6, but for the coronagraph images of the CME.}
\label{fitcor1}
\end{figure*}

The free parameters were varied until we found a satisfactory
projection of the model on the plane of the sky for {\it both} STA and STB.
Only the images taken from 18:41~UT onward could be fit because it was
hard to identify the erupting bubble with confidence at earlier times.
We emphasize that the simultaneous fitting for STA and STB places strong
constraints on the free parameters of the model. For instance, we found
that deviations from the best-fit values lead to non-tolerable solutions
already if the distance to the midpoint is changed by more than
$\approx\!5\%$ or the radius is changed by more than $\approx\!15\%$.
Error bars were obtained by such systematic variation of the fit
parameters. Note that these are asymmetric for the radius, i.e., the
tolerable values for its increase and reduction differ.

The fits yield a tilt of the CME leg by $\approx\!35^\circ$ with
respect to the radial direction, which is only weakly changing across
the analyzed height range. To our knowledge, this is the first time
that such strong non-radial evolution is  quantified so low in the corona. 
CMEs are often observed to start more radially in EUV images but can
become deflected towards the equator at larger heights,  especially
at solar minimum (e.g., Plunkett et al. 1997).
Non-radial motion should be given more attention in future CME modeling.

Due to the tilt, the distances to the front and midpoint of the fitted 
sphere differ from the corresponding heights above the photosphere by a
(nearly uniform) factor of $\approx\!1.2$. Since we are interested in the 
evolution of the CME flux rope, i.e., how it is accelerated along its 
propagation path and how it expands in relation to this acceleration, 
we will base the subsequent analysis on the distances along the
inclined propagation path. Since the difference to the true radial
heights is small and in order to 
conform to the usual designation in describing CMEs, we will simply refer
to these distances as ``heights'' in the remainder of the paper.

Figures \ref{fitb} and \ref{fitcor1}  show our fits for several
representative snapshots in the course of the event from EUVI and COR1,
respectively. The model reproduces the position and envelope
of the bubble quite well at all times and simultaneously for both
spacecraft, which justifies its use
in deriving quantitative information about the early evolution of the CME.
The fit is less satisfactory in the conical section of the employed
model. For example, the lowest parts of the observed bubble show a
stronger divergence with height. Also, the internal structure of the
bubble is not included. However, the development of a more
sophisticated model which can improve on these aspects is beyond the
scope of the present paper.

The 3D geometrical modeling gives us rather accurate values for the
height $h$ of the bubble center (along the inclined propagation path)
and for the radius $r$ of the bubble vs. time for further analysis.
These are independent of the quality of fit by the conical section of
the model. The rise profiles of the front edge and midpoint of the
bubble, $h(t)+r(t)$ and $h(t)$, respectively, and the radial evolution
$r(t)$ are plotted in Figs.~\ref{fitht} and \ref{meas} and are analyzed
in the following section.

\section{Timing of bubble expansion and acceleration, and relation to
         the associated flare}
\label{timing}

To characterize the main acceleration phase of the ejection and compare
it to the signatures of the associated flare, we derive velocity and
acceleration profiles from the height-time data. Here we follow the
most widespread practice to use the position of the CME front edge,
$h(t)+r(t)$, to facilitate the comparison with other studies. A twofold
application of smoothing and finite differencing is often used for this
purpose, but is known to yield large and often non-acceptable scatter in
the derived acceleration profile, even if the main acceleration phase is
sampled at a higher rate than here. Therefore, we use a further fit,
adopting the following function \citep{Sheeley&al2007}
\begin{eqnarray}
H(t)=H(t_1) +\frac{1}{2}(\upsilon_{f} + \upsilon_{0})(t-t_1) + \nonumber \\ 
     \frac{1}{2}(\upsilon_{f}-\upsilon_{0})\tau 
     \ln\left[\cosh\left(\frac{t-t_1}{\tau}\right)\right], 
\label{e:fitht}
\end{eqnarray}
whose derivatives yield the velocity and acceleration profiles as
\begin{eqnarray}
\upsilon(t)&=&\frac{1}{2}(\upsilon_{0}+\upsilon_{f})+
              \frac{1}{2}(\upsilon_{0}-\upsilon_{f})\,
              \tanh\left(\frac{t-t_1}{\tau}\right),
\label{e:fitvt}\\ 
a(t)&=&\frac{\upsilon_{f}-\upsilon_{0}}{2\tau}
       \left[1-{\tanh}^{2}\left(\frac{t-t_1}{\tau}\right)\right].
\label{e:fitat}
\end{eqnarray}
Here, $\upsilon_{0}$ and $\upsilon_{f}$ are initial and final asymptotic
velocities, respectively, $t_1$ is the time of peak acceleration (equal
to the time when the velocity reaches its average value
$(\upsilon_{0}+\upsilon_{f})/2$, due to the symmetry of the function),
and $\tau$ is the time scale of the rise to peak acceleration.

\begin{figure}
\includegraphics[width=0.62\textwidth]{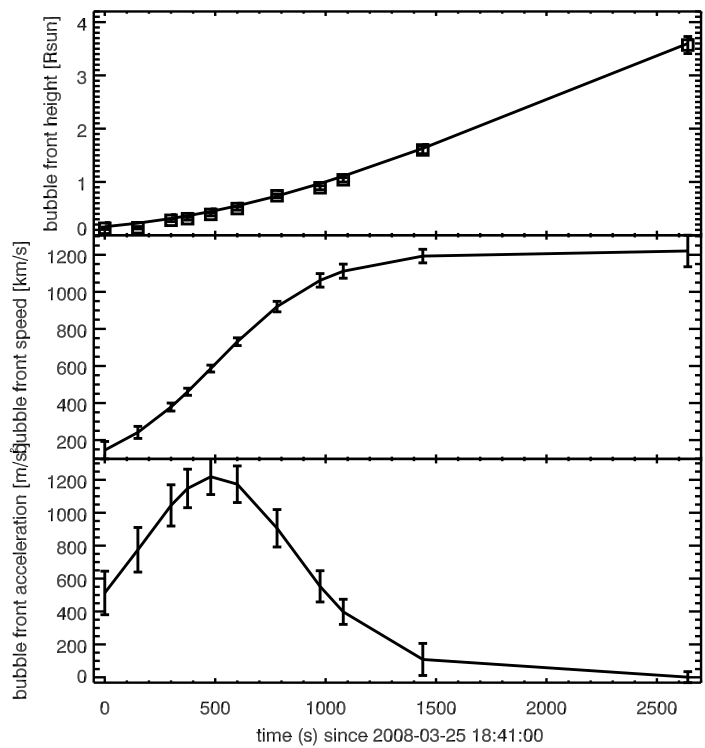}
\caption{Time evolution of bubble front height, $h(t)+r(t)$, from EUVI and
  COR1 data (final two datapoints) and its fit by the function given in
  Eq.~(\ref{e:fitht}) \emph{(upper panel)}.
  Velocity \emph{(middle)} and acceleration \emph{(bottom)} of bubble
  front derived from the fit. Error bars show $1\sigma$ uncertainties
  obtained by Monte Carlo simulations (see Sects. 4 and 5 for detail).}
\label{fitht}
\end{figure}

This function can reproduce profiles ranging from nearly constant
acceleration (when $\tau$ is comparable to or bigger than the considered
time interval) to impulsive acceleration (when $\tau$ is significantly 
smaller). It resembles shapes obtained in CME simulations
\cite[e.g.,][]{2006ApJ...644..592R, 2007AN....328..743T}, and is found
to fit our height data very well, as it did for slow streamer ejections
in \citet{Sheeley&al2007}. A slight drawback consists in the symmetry of
its acceleration profile, which likely influences the resulting time and
magnitude of the peak acceleration and the onset of fast acceleration
somewhat. Since the rise time of eruptions is often shorter than their
decay time, this fit function may place the onset slightly too early.
However, given the small number of our height data points (only eleven points), we do
not consider extended experimenting with the fit function warranted.

Uncertainties in the resulting velocity and acceleration profiles were
estimated by Monte Carlo simulation. We randomly perturbed the observed
height data within their estimated error bars and fitted these data with
Eq.~(\ref{e:fitht}). This proccess was repeated $10^4$ times and the
$1\sigma$ variations in the resulting velocity and acceleration profiles
are taken as estimates of their uncertainties.

The fit using Eq.~(\ref{e:fitht}) confirms the impulsive nature of the
acceleration profile (Fig.~\ref{fitht}) and permits us to estimate
its salient properties. The acceleration peaks at $t_1\approx500$~s
(18:49:20~UT) relatively low in the corona ($H(t_1)\approx0.4R_\odot$;
$h(t_1)\approx0.27R_\odot$)
and lasts for about 1000~s (FWHM). Nearly all of the acceleration occurs
within the EUVI field of view. The asymptotic velocity of
1200~$\mathrm{{km}\,{s}^{-1}}$, essentially reached at our first COR1 data
point at a heliocentric distance of the CME front of $\approx\!2.6R_\odot$, is
consistent with the velocity of $\approx\!1100$~km\,s$^{-1}$
obtained from a 3D fit to the COR2 data in Thernisien et al. (2009),
since fast CMEs generally decelerate slowly in the solar wind.
The fit indicates that the impulsive acceleration phase commenced around
18:36~UT.

\begin{figure}
\includegraphics[width=0.62\textwidth]{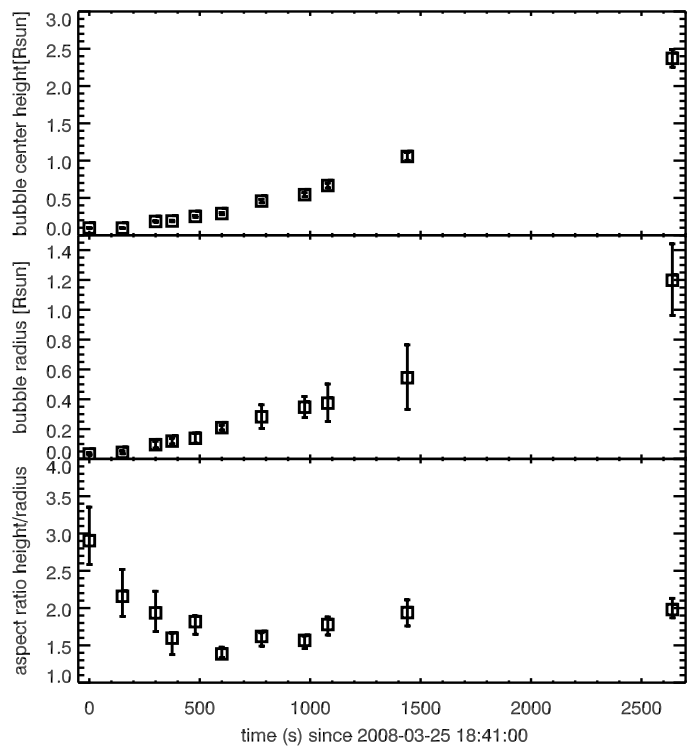}
\caption{Time evolution of bubble center height $h$, radius $r$,
  and aspect ratio $\kappa$,
  as determined by the 3D ice-cream cone model from simultaneous fitting of
  the STA and STB EUVI and COR1 observations.} 
\label{meas}
\end{figure}

In Fig.~\ref{meas} we plot the heights of the bubble center, $h(t)$,
joint with the corresponding
radii. The very small volume of the erupting flux prior to the onset of
the fast expansion is apparent: the first two data points yield bubble
radii $r<0.05R_\odot$. Next we combine these values to obtain the
aspect ratio of the bubble, $\kappa=h/r$. The error bars given
for this quantity in Fig.~\ref{meas} were calculated as the
root mean squared (rms)
values of all possible combinations of $h$ and $r$ errors (two for the
upper limit and two for the lower limit) for each data point. Standard
error propagation could not be used due to the asymmetry of the $r$
errors.

The aspect ratio of the bubble exhibits a clear two-phase evolution. It
decreases rapidly in the first $\approx\!600$~s of our data, followed
by a very gradual recovery not very different from a
plateau. This confirms the impression obtained from the animated
EUVI images as described in Sect.~\ref{over}. \emph{The expansion of the
erupting flux in the main acceleration phase of the ejection thus
reveals two major effects of different duration, i.e., two phases.}
In addition to the conventional impulsive main upward acceleration of a
fast CME, the event exhibits an initial lateral overexpansion (a rapidly
decreasing aspect ratio) that coincides approximately with the rise to
peak upward acceleration. Our data indicate very clearly that the
overexpansion is present from the beginning of the main acceleration
phase and show that it ends earlier than the main acceleration. Whether the
overexpansion generally ends near peak acceleration must be clarified by
studying further events and the underlying physics.

The very slow subsequent evolution of the aspect ratio $\kappa(t)$
means that the flux rope
evolves approximately self-similarly. This behavior, known to be
typical for CMEs in their propagation phase after the main acceleration
\cite[e.g.,][]{2001ApJ...562.1045K, 2006ApJ...652..763T,
2009SoPh..256..111T}, is here found to commence
very early, already in the course of the main aceleration.

To our knowledge, the possible existence of two phases in the course of
a CME's main acceleration has not been observed previously.  In particular,
the fitting of a similar spherical model to the same EUVI data has led
\citet{2009AnGeo..27.3275A} to conclude that the bubble evolved
self-similarly throughout the main acceleration phase with an aspect
ratio $\kappa\approx1$. It appears that one major reason for the
different result lies in his choice of too large bubble radii at most
times in the main acceleration phase (which are not consistent with the
STB data; see Sect.~\ref{proper}).
When the fitted sphere extends approximately between
the front of the bubble and the coronal base, then $h\approx r$ and
$\kappa\approx1$ at all times. The other major reason
derives from his assumption that both $h(t)$ and $r(t)$ data
can be represented by fit functions which are valid \emph{throughout}
the main acceleration phase. This masks any possible two-phase
evolution. Moreover, the acceleration was assumed to be uniform
(commencing as a step function) in both $h$ and $r$ directions, which is
an inadequate representation of impulsive CME acceleration
(and introduces a tendency to infer a delayed onset time, found to be
18:38~UT).

The aspect ratio of erupting flux was also considered in
\citet{2001ApJ...562.1045K} for a sample of mostly slow CMEs at
heliocentric distances (2--30)~$R_\odot$, where approximately constant
values in the range $\approx\!(1\mbox{--}2.5)$ were found. Note that
\citet{2001ApJ...562.1045K} used the heliocentric distance and the
diameter of the cavity to define an aspect ratio, which can be
expressed, using our aspect ratio $\kappa$, as
$\Lambda=\kappa+(R_{\odot}-h)/2r$. For $h\gg R_\odot$ the quantities
differ by a factor 2 ($\Lambda\approx\kappa/2$). For $h\ll R_\odot$ one has
$\Lambda\approx R_\odot/2r$, thus the relationship to $\kappa$ is
essentially lost in the height range where fast CMEs often commence.
\citet{2001ApJ...562.1045K} plot $\Lambda$ values also at low heights,
estimated from EUV images, for several events in their sample. However,
it is impossible to transform these data into the corresponding values
of $\kappa$ because the corresponding heights are not given with the
required high accuracy. Moreover, these data did not cover the relevant
height range of strongly rising acceleration.

\begin{figure}
\includegraphics[width=0.50\textwidth]{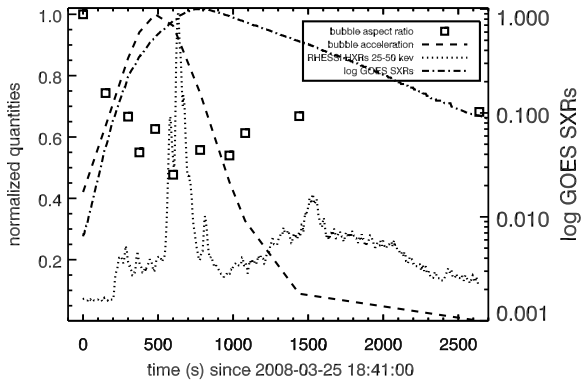}
 \caption{Evolution of bubble aspect ratio (squares, from
  Fig.~\ref{meas}) and acceleration (dashed, from Fig.~\ref{fitht}) 
  compared to the 25--50~keV light curve (dotted) from \textsl{RHESSI}
  \citep{2002SoPh..210....3L} and to the 1--8~\AA\ light curve
  (dashed-dotted) from \textsl{GOES} \citep{1994SoPh..154..275G}.
  Normalized quantities are plotted. Note that \textsl{RHESSI} was in
  spacecraft night prior to 18:45~UT ($t<240$~s).}
\label{goes}
\end{figure}

The initial value of the aspect
ratio in our data, $\kappa\approx3$ at 18:41~UT, suggests that a flux
rope was formed by this time and that it was detached from the
photosphere \cite[as opposed to a bald-patch topology; e.g.,][]{Titov&al1993}.
This suggests the existence of a current sheet beneath the flux rope,
where reconnection could produce flare signatures and add flux to the
rising rope, consistent with the strong rise of the flare emissions
after 18:41~UT.

A closer look at the aspect ratio after the phase of overexpansion
indicates a very gradual increase. To check whether this trend persists,
we compare our aspect ratios to the corresponding values obtained by
\citet{2009SoPh..256..111T} in the COR2 range. The full Graduated
Cylindrical Shell model is more appropriate than the ice-cream cone
model at these heights and yields a minor torus radius $r=2.4R_\odot$ at
the front height of $h+r=8.9R_\odot$ from an image pair near 20:22~UT
(this includes a moderate correction of the ratio $r/(h+r)$ quoted in
\citeauthor{2009SoPh..256..111T} [\citeyear{2009SoPh..256..111T}]). The
resulting aspect ratio $\kappa=2.7$ supports the trend indicated by our
spherical fits up to front heights $\lesssim4R_\odot$. However, it must
be noted that the proper definition of the aspect ratio may change in
the considered height range. Low in the corona, where we find the
overexpansion, the overall structure of the erupting flux is close to a
partial torus. Our choice of $\kappa$, center height by radius of the
bubble, corresponds to the aspect ratio of a torus (major by minor
radius) in this range. At large heights, $h\gtrsim R_\odot$ and beyond, 
the erupting
flux forms a nearly complete torus. The toroidal aspect ratio is then
better approximated by $\kappa/2$, which leads to an aspect ratio of
$\approx\!1.4$ at the front height of $8.9R_\odot$. 
Adopting values intermediate between $\kappa$ and $\kappa/2$
for our final two data points at $h>R_\odot$, the trend in
the COR1--COR2 height range is actually closer to flat
behavior or to a very gradual decrease of the aspect ratio.
This is consistent with the trend
found for the majority of events in \citet{2001ApJ...562.1045K}.
Although in this view the aspect ratio of the erupting flux may decrease
throughout the EUVI--COR2 height range, the existence of two clearly
distinguishable phases---rapid initial overexpansion, followed by very
gradual, approximately self-similar evolution---remains a robust result
for the event investigated here.

Figure~\ref{goes} displays the temporal relation between the
acceleration and aspect ratio of the CME cavity and the soft and hard
X-ray light curves of the associated flare. Previous studies of fast CMEs
have demonstrated a very close association between the CME acceleration
profile and the rise phase of the soft X-ray light curve in the majority
($\sim$\,50\%) of events, a moderately close association for a further
large fraction ($\sim$\,25\%), and substantial differences in the
remaining cases \citep{2007SoPh..241...99M}. The event studied here
falls in the second category: the acceleration profile peaks nicely on
the flank of the soft X-ray light curve, but slightly before the time of
steepest rise which coincides with the major hard X-ray pulse (this is
seen when the soft X-ray light curve is plotted on a linear scale as well).
Also, the acceleration commences earlier, by about 5~minutes, than the
main (exponential) rise of the soft X-rays, which, however, is not
uncommon. The onset of the acceleration is simultaneous to the onset of
the gradual rise of the soft X-ray emission above the base level set by
the precursor event in the neighboring active region.

Overall, the event shows a relatively high CME-flare correlation, as
the majority of fast CMEs does. An unusual role for reconnection is
not indicated. Consequently, if the overexpansion is related
to reconnection, we can expect that it occurs in many further events.
The strongest hard X-ray peak commences when the rapid
decrease of the bubble's aspect ratio ends, clearly indicating that the
overexpansion does not result from an enhanced reconnection rate,
rather it reflects an evolutionary stage of the eruption.

\section{Interpretation of bubble overexpansion}\label{interpretation}

The phase of overexpansion encompasses only a tiny fraction of the ascent
of the ejected flux (the center of the bubble rises from
$h\approx\!0.1R_\odot$ only to $h\approx\!0.29R_\odot$ in these
$\approx\!600$~s). Hence, it cannot primarily be due to the rise into an
environment of lower pressure. This effect would remain strong over a
much larger height range. Rather, a magnetic origin is implied.

Shearing and twisting the coronal field by
photospheric footpoint motions are known to produce huge inflations
\cite[e.g.,][]{1994ApJ...430..898M, Torok&Kliem2003}, but operate on far
longer time scales relevant to the energy storage phase and do not
exhibit a sudden onset, so these processes can also be excluded.

We suggest that one or a combination of the following two effects
causes the overexpansion. First, there is a purely ideal MHD effect of
the decreasing current in a rising flux rope. The decrease of the
current is a consequence of flux conservation between the flux rope and
the photospheric boundary \cite[e.g.,][]{Isenberg&Forbes2007}. It can be
made plausible by the fact that the number of field line turns in the
flux rope does not change under line-tied, ideal conditions when the
flux rope rises. This is equivalent to a decreasing azimuthal
(poloidal) field component and causes the flux surfaces to move apart
all around the rope. The weaker field is compensated by the larger
volume to conserve the flux between the surfaces, as required by ideal
MHD. Nominally, the expansion progresses outward at the Alfv\'en speed,
but the tension of the surrounding field resists the expansion, slowing
it down. Nevertheless, a cavity expands rapidly. Note that this
expansion extends into the volume \emph{outside} of the actual flux rope
and is a natural consequence of a preexisting flux rope.

\begin{figure*}
\includegraphics[]{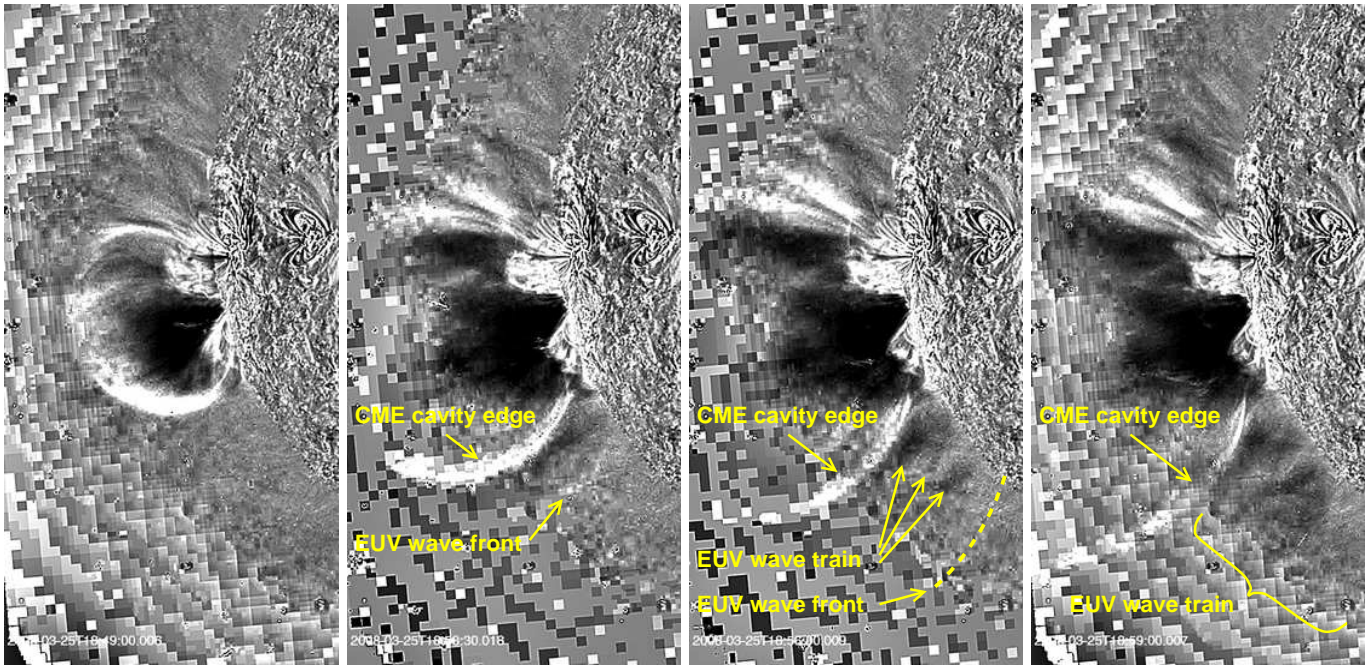}
\caption{Sequence of STA 171~{\AA} base-ratio images at 18:49, 18:53:30,
  18:56, and 18:59~UT, showing the launch of the EUV wave by the horizontal
  expansion of the CME cavity (which is far stronger expressed in
  southward direction, due to the initially nonradial rise of the CME)
  and the subsequent detachment of the wave front from the cavity.
  Due to the upward rise of the cavity, its horizontal expansion at heights
  $h\lesssim(0.2\mbox{--}0.3)R_\odot$ in both northern and southern
  directions has terminated by the time of the final panel.
  The wave in the northern direction is much slower, but can be seen
  clearly in the corresponding animations in the online edition.}
\label{f:wave}
\end{figure*}

Second, the overexpansion can also result from the rapid addition of
flux to the CME flux rope by magnetic reconnection in the vertical
(flare) current sheet underneath. In this case, the flux rope comprises the
whole observed cavity, and the overexpansion can signify the growth of a
preexisting flux rope \citep{2000JGR...105.2375L}, as well as the
formation of the flux rope from arcade field lines
\citep{2008ApJ...683.1192L}.

Numerical modeling of overexpanding rising flux ropes,
planned for a follow-up investigation, is expected to reveal which of
these effects conforms best to the data analyzed here. Also, it should
show whether the nearly simultaneous end of overexpansion and of rising
acceleration in the studied CME represent a systematic effect or a
coincidence.

\section{Launch of the EUV wave}\label{s:wave}

Finally, let us briefly address the EUV wave which was observed both
off-limb and on-disk in association with the eruption. The launch of the
wave and its association with the rapidly expanding CME cavity is
clearly displayed in both base ratio movies provided in the online
edition. The wave is first seen ahead of the south rim in the 171 STA
frame at 18:49:45~UT and in the 171 STB frame at 18:50~UT.
Snapshots of the wave are also shown in Fig.~\ref{f:wave}. 

The STA base-ratio movie (see also Fig.~\ref{ratio} for individual frames)
shows that, at any
given height, the lateral expansion of the developing CME cavity has the
temporal profile of a single pulse. The lateral expansion starts
impulsively (as does the whole CME) but also ends rather abruptly, which
is due to the rise of the cavity and its subsequent detachment from the
solar surface. The wave propagates away from the cavity rim as soon as
the lateral expansion of the rim has begun to slow down at low heights.
These effects are more pronounced at the southern side of the eruption,
due to the southward inclination of the initial rise. The off-limb wave
front can be tracked by following the outermost deflected off-limb
structures. Their locations roughly coincide with the latitudinal extent
of the wave on the disk.

The decoupling between the erupting flux and the EUV wave shows
that the observed intensity front is a freely travelling wave.
This is at variance with suggestions that EUV waves represent
the lower coronal extension of the CME---a ``current shell'' at the
surface of the expanding flux \citep{2008SoPh..247..123D}, or the
``magnetic footprint'' of a CME ``skirt'' which reconnects with the
ambient quiet-Sun flux \citep{2007ApJ...656L.101A}. This has already
previously been found, based on 3D fittings of EUVI and COR1 images
of two slow CMEs with associated EUV waves \citep{Patsourakos_09sol,
2009ApJ...700L.182P}.
The animated base-ratio images of the present event and our 3D
fitting of the expanding flux at the relevant low heights provide
particularly clear support for the interpretation of EUV waves as freely
traveling waves. In addition, they show that it is the impulsive
horizontal expansion of the fast CME low in the corona which triggers
the wave.

\section{Summary and conclusions}\label{concl}

We present a detailed analysis of the formation and early
evolution phase of an impulsively accelerated, fast CME, beginning to
exploit the diagnostic potential of its expanding cavity. 
Key aspects of this work are the 3D modeling of CME expansion, based
on two-viewpoint observations, and the continuous coverage of the
critical range of CME formation and main acceleration in the inner
corona, and up to several solar radii by the \textsl{STEREO} mission.
These allow us to quantify the initial expansion of the erupting
flux to obtain new insight into the genesis of CMEs.
The observational results can be summarized as follows:
\begin{itemize}
\item The CME forms \emph{a rapidly expanding ``bubble''} simultaneously
  with the onset of the main upward acceleration. 
\item The bubble appears as a nearly circular area of low EUV emission
  surrounded by a relatively narrow and bright rim but is a
  three-dimensional structure that can be described reasonably well by a
  sphere in its early development stages.
\item The bubble evolves into the cavity of the white-light CME, which
  is well fit by a geometric flux rope model at distances
  $>2.5~R_{\odot}$ (Thernisien et al. 2009). Hence, it is highly likely
  that \emph{the bubble is the early signature of the flux rope}
  commonly suggested by CME models.
  The associated erupting prominence rises with the bottom part of
  the bubble.
\item The aspect ratio of the bubble, center height vs.\ radius, evolves
  non-linearly in \emph{two phases}. An initial rapid decrease, which
  signifies an overexpansion faster than the rise, is followed by
  nearly constant aspect ratio, i.e., a nearly self-similar evolution.
\item The overexpansion is present from the beginning of the
  formation of the bubble but ends earlier than the main upward
  acceleration, close to the acceleration peak.
\item The main acceleration is preceded by a slow-rise phase of
  active-region loops overlying the prominence. The images in this phase
  are inconclusive with regard to the existence (or absence) of a cavity
  enclosed by these loops.
\item The explosive nature of the eruption induces deflections
  of ambient coronal structures and launches an EUV wave propagating
  across the solar disk. The EUV wave is likely triggered by the
  pulse-like horizontal component of CME cavity (bubble) expansion
  low in the corona. As the horizontal expansion in the low corona
  slows down and terminates, due to the further rise of the erupting
  flux, the EUV wave becomes a freely propagating wave.
\item As for most fast CMEs, the main upward acceleration is relatively
  well synchronized with the impulsive rise phase of the associated
  flare soft X-ray emission.
\end{itemize}

We exclude that the initial overexpansion of the erupting flux is
caused by decreasing ambient pressure (as the flux rises) or by
photospheric motions and suggest that
it results from one or a combination of the following two effects.
First, an expansion of the flux surfaces of the poloidal flux external
to the actual rope, which is caused by flux conservation for
decreasing current through the rope. The decrease of the current results
from the rise of a line-tied rope in ideal MHD. Second,
the addition of flux by reconnection in a current sheet
under a strongly developing, possibly newly forming, flux rope.

Our main conclusions are the following.
\begin{enumerate}
\item The three-dimensional expansion of the investigated fast CME
  exhibits two phases in the course of the main upward acceleration.
\item The first phase is characterized by an overexpansion, i.e.,
  decreasing aspect ratio, of the CME cavity.
\item The second phase is characterized by approximately self-similar
  evolution of the cavity.
\item The pulse-like horizontal expansion (parallel to the solar
  surface) of the CME cavity at low coronal heights triggers a freely
  propagating EUV wave.
\end{enumerate}

Being obtained for a single event, these results refer explicitly to
impulsively accelerated, fast CMEs. They require substantiation through
the study of further eruptions, which is currently underway. Since fast
and slow CMEs may eventually be explained by a single physical model
\cite[e.g.,][]{Vrsnak&al2005, 2007AN....328..743T}, our results should
be relevant for CMEs in general.

\begin{acknowledgements}
The SECCHI data used here were produced by an international consortium
of the Naval Research Laboratory (USA), Lockheed Martin Solar and
Astrophysics Lab (USA), NASA Goddard Space Flight Center (USA),
Rutherford Appleton Laboratory (UK), University of Birmingham (UK),
Max$-$Planck$-$Institut for Solar System Research (Germany), Centre
Spatiale de Li\`ege (Belgium), Institut d'Optique Th\'eorique et
Applique\'e (France), and Institut d'Astrophysique Spatiale (France).
We gratefully acknowledge very constructive discussions with
T.~G.~Forbes about cavity expansion, which significantly helped us to
interpret the results. We also thank the referee and Hardi Peter for helpful comments.
This work was suported by NASA grants NNH06AD58I and NNX08AG44G,
by an STFC Rolling Grant, and by the DFG.
\end{acknowledgements}

\end{document}